\documentclass[aps,pra,twocolumn,superscriptaddress]{revtex4-1}

\usepackage{soul}
\usepackage{relsize}
\usepackage{lmodern}
\usepackage{slantsc}
\usepackage{graphicx}
\usepackage{amsmath}
\usepackage{amssymb}
\usepackage{amsthm}
\usepackage{amsbsy}
\usepackage{mathrsfs}
\usepackage{varioref}
\usepackage{dsfont}
\usepackage{bm}
\usepackage{color}
\usepackage[usenames,dvipsnames]{xcolor}
\definecolor{myblue}{rgb}{0.2,0.2,0.8}
\definecolor{myblack}{rgb}{0,0,0}
\definecolor{myurl}{rgb}{0.1,0.1,0.4}
\usepackage[colorlinks=true,citecolor=myblue,linkcolor=myblack,urlcolor=myurl,hyperindex,breaklinks]{hyperref}
\usepackage{cleveref}
\usepackage{subfigure}
\usepackage{booktabs} 
\usepackage{array} 
\usepackage{paralist} 
\usepackage{verbatim} 
\edef\restoreparindent{\parindent=\the\parindent\relax}
\usepackage[parfill]{parskip}
\restoreparindent
\usepackage{natbib}

\setul{}{20 pt}

\newcommand{\<}{\langle}
\renewcommand{\>}{\rangle}

\newcommand{\ww}{{\mathcal{W}}}

\newcommand{\lo}{{\mathcal{L}}}

\newcommand{\trc}{{\mathcal{T}}}
\newcommand{\s}{{\mathcal{S}}}

\renewcommand{\aa}{{\mathcal{A}}}

\newcommand{\mm}{{\mathcal{M}}}

\newcommand{\ii}{{\mathcal{I}}}

\newcommand{\jj}{{\mathcal{J}}}
\newcommand{\co}{\mathds{C}}

\newcommand{\h}{{\mathcal{H}}}

\newcommand{\ha}{{\mathcal{H}\sub{\aa}}}

\newcommand{\one}{\mathds{1}}
\newcommand{\zero}{\mathds{O}}

\newcommand{\tr}{\mathrm{tr}}

\newcommand{\sub}[1]{_{\!\mathsmaller{#1}}}
\newcommand{\sups}[1]{^{\mathstrut{\mathsmaller{#1}}}}
\newcommand{\eq}[1]{Eq.~\eqref{#1}}

\newcommand{\sect}[1]{Sec.~\ref{#1}}
\newcommand{\app}[1]{Appendix~(\ref{#1})}
\newcommand{\ket}[1]{|{#1}\rangle}

\newcommand{\pr}[1]{P[{#1}]}

\newcommand{\avg}[1]{\langle {#1} \rangle}

\begin{document}

\title{Self-consistency of the two-point energy measurement protocol}

\author{M. Hamed Mohammady}
\email{m.hamed.mohammady@savba.sk }
\affiliation{RCQI, Institute of Physics, Slovak Academy of Sciences, D\'ubravsk\'a cesta 9, Bratislava 84511, Slovakia}


\begin{abstract}
A thermally isolated quantum system undergoes unitary evolution by interacting with an external work source. The two-point energy measurement (TPM) protocol defines the work exchanged between the system and the work source by performing ideal energy measurements on the system before, and after, the  unitary evolution.  However,  the ideal energy measurements used in the TPM protocol ultimately result from a unitary coupling with a measurement apparatus, which  requires an interaction with an external work source. For the TPM protocol to be self-consistent, we must be able to perform the TPM protocol on the compound of system plus apparatus, thus revealing the total work distribution, such that when ignoring the apparatus degrees of freedom, we recover the original TPM work distribution for the system of interest. In the present manuscript, we show that such self-consistency is satisfied so long as the apparatus is initially prepared in an energy eigenstate. Moreover, we demonstrate that if the apparatus Hamiltonian is equivalent to the ``pointer observable", then: (i)  the total work distribution will satisfy the ``unmeasured'' first law of thermodynamics for all system states and system-only unitary processes;  and (ii) the total work distribution will be identical to the system-only work distribution, for all system states and system-only unitary processes, if and only if the  unmeasured work due to the unitary coupling between system and apparatus is zero for all system states.
\end{abstract}

\maketitle

\section{ Introduction}

The definition of work for quantum systems is one of the most contentious issues in quantum thermodynamics, and continues to be a subject of heated debate \cite{Allahverdyan2005, Talkner2007, Skrzypczyk2014b, Watanabe2014,  Roncaglia2014, Halpern2014, Gemmer2015, Gallego2016, Hayashi2017, Faist2017, Niedenzu2019, Sone2020}. The paradigmatic scenario is the work done on a thermally isolated system: a system which  is only mechanically manipulated, by means of inducing time-dependence on its Hamiltonian,  and thus evolves unitarily. Such mechanical manipulation results from an interaction with an external work source, and is hence generally accompanied by an exchange of work. In the limiting case where the system starts and ends in a classical mixture of energy eigenstates,   in any given realization the work done on the system is well defined, and is the difference in energy eigenvalues. By performing ideal energy measurements before, and after, the unitary evolution, one can therefore observe which particular value of work obtains in any given realization without disturbing the system. Furthermore, the average work done, given by the observed probability distribution over work, will  be equivalent to the difference in average energies evaluated before, and after, the unitary evolution; the  ``unmeasured'' first law of thermodynamics is satisfied. The two-point energy measurement (TPM) protocol extends this procedure for determining the work distribution, namely, performing ideal energy measurements before and after the unitary evolution, to general unitary processes and general states \cite{Esposito2009,Campisi2011}.  
However,  in general if the initial state does not commute with the Hamiltonian, the unmeasured first law will be violated; the average work obtained by the TPM protocol will  not coincide with the difference in average energies. Indeed, as shown in Ref. \cite{Perarnau-Llobet2016a}, no measurement procedure exists which simultaneously recovers the  work distribution for systems   in a classical mixture of energy eigenstates, and recovers the average work as the unmeasured work,  for all states and unitary processes.

That the  TPM protocol cannot always satisfy the unmeasured first law ultimately rests on one of the central maxims of quantum measurement theory: no information without disturbance \cite{Busch2009}. To be sure, ideal measurements are the least disturbing measurements available \cite{Lahti1991,Busch1995},  but only insofar as there are \emph{some} states that are undisturbed by such measurements. This perceived failure of the TPM definition has lead to alternative formulations of work, such as defining work as the unmeasured work  \emph{simpliciter} \cite{ Anders2013,  Deffner2016a,Strasberg2019}, and the Margenau-Hill method and related approaches using quasi-probability distributions \cite{Marg-Hill, Allahverdyan2014, Miller2016, Lostaglio2018, Diaz2020}. 

Of course, there is another issue raised by the TPM protocol, or indeed any method which uses measurement as part of the definition for work: can such a method be self-consistent? The ideal energy measurements used in the TPM protocol must ultimately result from a \emph{physical} interaction between the system and a measurement apparatus. The quantum theory of measurement allows for the  measurement of any observable to be physically modeled as a \emph{normal measurement scheme}, which involves a  unitary  interaction between the system and a measurement apparatus which is initially prepared in a fixed pure state -- a condition which is possible to satisfy in principle,  thermodynamic limitations on preparing pure states notwithstanding \cite{Masanes2014, Debarba2019, Mohammady2019c, Guryanova2018} -- followed by measurement of the apparatus by a sharp \emph{pointer observable} \cite{Ozawa1984}. Normal measurement schemes have been used to ``indirectly'' measure work \cite{Roncaglia2014, DeChiara2018}. Of course, such unitary interactions between the system and apparatus themselves result from mechanically manipulating this compound system, and hence are generally  accompanied with an exchange of work with an external work source. If the TPM definition of work is valid for the system of interest, therefore, it stands to reason that it is valid for the compound of system plus apparatus; by performing ideal energy measurements on both system and apparatus, before and after the total unitary evolution of the compound system, we thus obtain the total work distribution.  We shall say that a given measurement scheme for the TPM protocol is self-consistent if the marginal work distribution for the system, obtained  when ignoring the apparatus degrees of freedom, is identical to the original system-only TPM work distribution, for all system states and system-only unitary processes. This idea is inspired by the theory of quantum incompatibility, where two observables are said to be compatible if there exists a third observable, the marginals of which recover the original observables in question \cite{Heinosaari2015}. In the present manuscript, we show that such self-consistency is always achieved if the apparatus is initially prepared in an energy eigenstate. 

 Interestingly, if we further restrict the measurement scheme such that the apparatus Hamiltonian is equivalent to the pointer observable used to measure the apparatus, then the total work distribution will always satisfy the unmeasured first law; the average total work will be the difference in average energy given the total unitary evolution, for all system states and system-only unitary processes (i.e. excluding the apparatus state, and the unitary interaction between system  and apparatus, which are fixed by the chosen measurement scheme). This is a consequence of the strong repeatability of ideal energy measurements \cite{Busch1995}, which implies that given the unitary interaction between system and apparatus, followed by measurement of the apparatus by the pointer observable, ``directly'' performing an ideal energy measurement on the system is superfluous. Of course, this statement should not be taken as a refutation of Ref. \cite{Perarnau-Llobet2016a}, since the initial state of the apparatus is always fixed, and commutes with the Hamiltonian by construction. But this observation does illustrate that it is possible for the  TPM work distribution to satisfy the unmeasured first law  for a large class of  initial states that do not commute with the Hamiltonian. 
 
 Finally, in the case where the apparatus is initially prepared in an energy eigenstate, and the  apparatus Hamiltonian is equivalent to the pointer observable,  we show that the total work distribution will be identical with the system-only work distribution, for all system states and system-only unitary processes,  if and only if the subspace of the apparatus which is involved during the measurement process corresponds with a single degenerate subspace of the apparatus Hamiltonian. This condition is further shown to be equivalent to the statement that the unmeasured work, due to the unitary interaction between system and apparatus, vanishes for all system states.

\section{TPM protocol}\label{sec:TPM}

We consider systems with a separable Hilbert space $\h$, with $\lo(\h)$ the algebra of bounded operators on $\h$, $\trc(\h)\subseteq \lo(\h)$ the space of 
trace-class operators, and   $\s(\h)\subset \trc(\h)$ the space of positive unit-trace operators (states), respectively. Moreover, we shall assume that the system is  thermally isolated,  with a bounded, time-dependent  Hamiltonian $H(t) = H + H_I(t)$. Here, $H$ is the system's ``bare'' Hamiltonian, describing it when it is fully isolated, i.e., isolated both thermally and mechanically. We assume this Hamiltonian to have a discrete spectrum, and may thus write it as 
\begin{align}
&H = \sum_{m} \epsilon_{m} P_m.
\end{align}
Here,  $\epsilon_{m}$ are  energy eigenvalues, and $P_m \geqslant \zero$ the corresponding spectral 
projections such that $P_m P_n = \delta_{m,n} P_m$ and  $\sum_{m} P_m = \one$. By the spectral theorem, the bare Hamiltonian $H$ is associated with a discrete, sharp observable $P := \{P_m \}$, where $m$ are the measurement outcomes which, given a state preparation $\rho \in \s(\h)$, are observed with the probability $\tr[P_m \rho]$ \cite{Heinosaari2011}. 

The time-dependence of $H(t)$ is entirely due to the term $H_I(t)$, which results from mechanically coupling the system with an external work source. If we assume that the system is only coupled with the 
work source for times $t \in (t_0, t_1)$, such that  $H_I(t) = \zero$ for all $t \leqslant t_0$ and $t \geqslant t_1$, then  the system's time 
evolution due to its interaction with the work source will be  described by the unitary  operator    $V := 
\overleftarrow{T} \exp(-i\int_{t_0}^{t_1} dt \, H(t))$, where we note that throughout this manuscript we use $\hbar = 1$  \cite{Allahverdyan2014}.   Given an initial state preparation $\rho$, the unmeasured work is thus 
\begin{align}\label{eq:unmeasured-work}
W := \tr[(V^\dagger H V - H) \rho].
\end{align}

The TPM protocol, for revealing the distribution of work due to the interaction between the system and the work source, is  given by the following sequence of operations:
\begin{enumerate}[(i)]
\item At time $t=t_0$, perform an ideal measurement of the bare Hamiltonian on the system, which is initially in an arbitrary state $\rho$. Given that outcome $m$ is observed, the system will be prepared in the (unnormalized) state 
\begin{align}
P_m \rho P_m.
\end{align}

\item Between time $t_0$ and $t_1$, let the system evolve unitarily, given its interaction with the external work source. The system will thus be prepared in the (unnormalized) state 
\begin{align}
 V P_m \rho P_m V^\dagger.
 \end{align}

\item At time $t=t_1$, perform an ideal measurement of the bare Hamiltonian on the system. Given that outcome $n$ is observed, the system will be prepared in the (unnormalized) state 
\begin{align}\label{eq:instrument-TPM-final-unnormalized-state}
P_n V P_m \rho P_m V^\dagger P_n.
\end{align} 

\end{enumerate}

The sequence of energy measurement outcomes $x:= (m,n)$ thus corresponds with the work done $w(x) := \epsilon_n - \epsilon_m$, and its probability is  given by the Born rule as the trace of the final unnormalized state \eq{eq:instrument-TPM-final-unnormalized-state}, which reads
\begin{align}\label{eq:TPM-probability}
p_\rho^V(x) := \tr[ P_m V^\dagger P_n V P_m \rho].  
\end{align}
Therefore the probability distribution for the work done, $w$, given the initial state $\rho$ and unitary operator $V$, is  
\begin{align}\label{eq:distribution-work}
p_\rho^V(w) := \sum_{x}\delta(w -w(x) ) p_\rho^V(x),
\end{align}
where $\delta(a - b) = 1$ if  $a=b$, and is zero otherwise. The average work can thus be computed to be    
\begin{align}\label{eq:TPM-average-work}
\avg{w}_\rho^V &:= \sum_w p^V_\rho(w) w \equiv \sum_{x} p_\rho^V(x) w(x), \nonumber \\
&= \tr[(\ii^L_\mm(V^\dagger H V) - H)\rho], \nonumber \\
& \equiv \tr[( V^\dagger H V - H)\ii^L_\mm(\rho)], 
\end{align}
where $\ii^L_\mm(\cdot) := \sum_{m} P_m (\cdot) P_m$ is the L\"uders channel for the bare Hamiltonian $H$. Given that for any $A \in \lo(\h)$, $\ii^L_\mm(A) = A$ if and only if $[H, A]=\zero$  \cite{Weihua2010},  it follows that  $\avg{w}_\rho^V = \tr[(V^\dagger H V - H) \rho]$ for all $V$  (for all $\rho$) only if $[H, \rho]=\zero$  ($[H, V^\dagger H V]=\zero$). In other words, the unmeasured first law \eq{eq:unmeasured-work} cannot be satisfied for all states and all unitary processes.

\subsection{Introducing the measurement apparatus in the TPM protocol}\label{sec:TPM-measurement-scheme} 
As shown above, the TPM protocol relies on performing ideal energy measurements on the system of interest both before, and after, the unitary evolution $V$. Such measurements are physically realized by an appropriate interaction between the system of interest and a measurement apparatus. The quantum theory of measurement allows all measurements on the system of interest to be modeled as a \emph{normal measurement scheme} \cite{Ozawa1984,PaulBuschMarianGrabowski1995}. Here,  the system of interest first interacts with a quantum ``probe'' of a measurement apparatus, initially prepared in a fixed pure state, by an appropriate unitary operator. Subsequently, the probe is measured by an appropriate pointer observable, and the measurement outcome observed indicates that the corresponding outcome has been observed for the desired system observable.

Since two energy measurements are performed on the system during the TPM protocol, we can generally consider the apparatus to be composed of two probes, one of which interacts with the system at time $t=t_0$, and the other at time $t=t_1$. As such, for the ideal energy measurement performed at time $t_j$, we may mathematically describe the normal measurement scheme by the tuple  $(\h\sub{\aa}^{(j)}, \ket{\xi^{(j)}}, U^{(j)}, Z^{(j)})$, where: $\h\sub{\aa}^{(j)}$ is the Hilbert space for the probe used, which is initially prepared in the pure state $\ket{\xi^{(j)}}$; $Z^{(j)} := \{Z^{(j)}_m\}$ is a sharp pointer observable, which has the same outcomes  as the system observable $P := \{P_m \}$; and $U^{(j)} $ is a joint unitary operator on the compound Hilbert space $\h \otimes \h\sub{\aa}^{(j)}$. This normal measurement scheme will realize an ideal measurement of the bare Hamiltonian on the system of interest if, for all $T \in \trc(\h)$ and $m$, we have 
\begin{align}\label{eq:Luders-dilation}
\tr_{\h\sub{\aa}\sups{(j)}}[(\one \otimes Z^{(j)}_m) U^{(j)}(T \otimes \pr{\xi^{(j)}})U^{(j)\dagger}] =P_m T P_m,
\end{align}
where $\pr{\xi^{(j)}} \equiv |\xi^{(j)}\>\<\xi^{(j)}|$ is a projection on the unit vector $\ket{\xi^{(j)}} \in \h\sub{\aa}^{(j)}$, and  $\tr_{\h\sub{\aa}\sups{(j)}} : \trc(\h\otimes \h\sub{\aa}^{(j)}) \to \trc(\h)$ is the partial trace  over the probe \cite{Busch1996,Busch2016a}.  It is simple to verify that in order for the  unitary $U^{(j)}$ to satisfy \eq{eq:Luders-dilation}, it must satisfy
\begin{align}\label{eq:Luders-unitary}
U^{(j)}(\ket{\psi} \otimes \ket{\xi^{(j)}}) = \sum_{m} P_m \ket{\psi} \otimes \ket{\phi_m^{(j)}}
\end{align} 
for all $\ket{\psi} \in \h$, where $\ket{\phi_m^{(j)}}$ are eigenvalue-1 eigenstates of the projection operators $Z^{(j)}_m$, i.e., $Z^{(j)}_n \ket{\phi_m^{(j)}} = \delta_{m,n}\ket{\phi_m^{(j)}}$ \cite{PeterMittelstaedt2004}. 

The TPM protocol can now be performed as follows:

\begin{enumerate}[(i)]

\item At time $t=t_0$, bring the system, initially prepared in an arbitrary state $\rho$, in contact with probe $\h\sub{\aa}^{(0)}$. The state of the compound system is thus $\rho \otimes \pr{\xi^{(0)}}$. Subsequently let the system interact with the probe  by the unitary operator $U^{(0)}$, which prepares the state 
\begin{align}
&U^{(0)}(\rho \otimes \pr{\xi^{(0)}}) U^{(0)\dagger}\nonumber \\
& \qquad  = \sum_{m,m'}P_m \rho P_{m'} \otimes |\phi_{m}^{(0)}\>\<\phi_{m'}^{(0)}|.
\end{align}
Finally, perform a measurement of the probe by the pointer observable $Z^{(0)}$. Given that outcome $m$ is observed, the  system will be prepared in the (unnormalized) state 
\begin{align}
P_m \rho P_m. 
\end{align}

\item Between time $t_0$ and $t_1$, let the system evolve unitarily, given its interaction with the external work source. The system will thus be prepared in the (unnormalized) state 
\begin{align}
V P_m \rho P_m V^\dagger.
\end{align}

\item At time $t=t_1$, bring the system in contact with probe $\h\sub{\aa}^{(1)}$. The (unnormalized) state of the compound system is thus $V P_m \rho P_m V^\dagger \otimes \pr{\xi^{(1)}}$. Subsequently let the system interact with the probe  by the unitary operator $U^{(1)}$, which prepares the (unnormalized) state 
\begin{align}
&U^{(1)}(V P_m \rho P_m V^\dagger \otimes \pr{\xi^{(1)}}) U^{(1)\dagger}\nonumber \\
& \qquad  = \sum_{n,n'}P_n V P_m \rho P_m V^\dagger P_{n'} \otimes |\phi_{n}^{(1)}\>\<\phi_{n'}^{(1)}|.
\end{align}
Finally, perform a measurement of the probe by the pointer observable $Z^{(1)}$. Given that outcome $n$ is observed, the system will be prepared in the (unnormalized) state 
\begin{align}\label{eq:TPM-measurement-scheme-final-state}
P_n V P_m \rho P_m V^\dagger P_n.
\end{align} 
\end{enumerate}
It is evident that the measurement scheme described above is identical to the original TPM protocol involving ``direct'' measurements on the system.

\section{ Consistently applying the TPM protocol to both system and apparatus}\label{sec:consistent-total-TPM}
The measurement scheme introduced in \sect{sec:TPM-measurement-scheme} does not make any assumptions regarding the Hamiltonian of the apparatus probes, nor the time it takes for the unitary operators $U^{(j)}$ to be generated; indeed, these were assumed to be implemented instantaneously. However, for the  unitary operator $U^{(j)}$ on $\h \otimes \h\sub{\aa}^{(j)}$ to be physical, it must also result from mechanically manipulating the Hamiltonian of this composite system, and thus requires an interaction with an external work source for a finite duration \cite{Strasberg2020a}. Let us therefore write the total time-dependent Hamiltonian  as $H_\mathrm{tot}(t) = H_\mathrm{tot} +  H_I(t) +  H_\mathrm{int}^{(0)}(t) + H_\mathrm{int}^{(1)}(t)$, where  $H_\mathrm{tot} = H + H\sub{\aa}^{(0)}  +  H\sub{\aa}^{(1)}$ is the additive, total bare Hamiltonian of system plus apparatus, and $H_I(t)$ is the system-only interaction Hamiltonian introduced in \sect{sec:TPM}. We shall denote the bare Hamiltonian of each probe in the spectral form as 
\begin{align}
H\sub{\aa}^{(j)} = \sum_\mu \lambda^{(j)}_\mu Q_\mu^{(j)},
\end{align}
where $\lambda_\mu^{(j)}$ are energy eigenvalues and $Q_\mu^{(j)}$ the spectral projections.  The interaction Hamiltonian for the composite system $\h \otimes \h\sub{\aa}^{(j)}$,   due to coupling with an external work source, is denoted $H_\mathrm{int}^{(j)}(t)$. Moreover,  $H_\mathrm{int}^{(0)}(t) = \zero$ for all $t \leqslant t_0'$ and $t \geqslant t_0$, and similarly $H_\mathrm{int}^{(1)}(t) = \zero$ for all $t \leqslant t_1$ and $t \geqslant t_1'$, where $t_0' < t_0 < t_1 < t_1'$. In other words, the interaction Hamiltonian $H_\mathrm{int}^{(0)}(t)$ is non-vanishing only for a finite duration before the system undergoes its isolated unitary evolution $V$, and similarly $H_\mathrm{int}^{(1)}(t)$ is non-vanishing only for a finite duration after the system undergoes its isolated unitary evolution $V$.  Therefore, by choosing the interaction Hamiltonians $H_\mathrm{int}^{(j)}(t)$ appropriately so that 
\begin{align}
 \overleftarrow{T} \exp\left(-i\int_{t_0'}^{t_0} dt \, \left[H + H\sub{\aa}^{(0)} + H_\mathrm{int}^{(0)}(t) \right] \right) = U^{(0)}, \nonumber \\
 \overleftarrow{T} \exp\left(-i\int_{t_1}^{t_1'} dt \, \left[ H +  H\sub{\aa}^{(1)} + H_\mathrm{int}^{(1)}(t) \right] \right) = U^{(1)},
 \end{align}
   the total unitary operator which describes the compound system's evolution during the extended period $t \in (t_0', t_1')$ will be 
\begin{align}\label{eq:total-unitary-system-apparatus} 
V_\mathrm{tot} &:= \overleftarrow{T} \exp\left(-i\int_{t_0'}^{t_1'} dt \, H_\mathrm{tot}(t)\right), \nonumber \\
&  = U^{(1)} (V \otimes e^{-i \theta_0 H\sub{\aa}^{(0)}} \otimes e^{-i \theta_1 H\sub{\aa}^{(1)}}) U^{(0)}, \nonumber \\
& \equiv  e^{-i \theta_0 H\sub{\aa}^{(0)}} U^{(1)} V U^{(0)} e^{- i \theta_1 H\sub{\aa}^{(1)}}.
\end{align}
Here, $e^{- i \theta_j H\sub{\aa}^{(j)}}$, where $\theta_0 = t_1' - t_0$ and $\theta_1 = t_1 - t_0'$, describes the contribution to the total unitary evolution from the bare Hamiltonian of probe $\h\sub{\aa}^{(j)}$, i.e., for the time period where the interaction Hamiltonian $H_\mathrm{int}^{(j)}(t)$ vanishes.  Note that the final line of \eq{eq:total-unitary-system-apparatus} is obtained because the unitary operators $e^{-i \theta_0 H\sub{\aa}^{(0)}}$ and $e^{-i \theta_1 H\sub{\aa}^{(1)}}$ commute with $U^{(1)}$  and $U^{(0)}$, respectively, since they act on different Hilbert spaces. Given an initial state preparation $\rho \in \s(\h)$, the total unmeasured work will thus read
\begin{align}\label{eq:total-unmeasured-work}
W_\mathrm{tot} := \tr[(V_\mathrm{tot}^\dagger H_\mathrm{tot} V_\mathrm{tot} - H_\mathrm{tot}) \rho \otimes \pr{\xi}],
\end{align}
where  $\ket{\xi}:= \ket{\xi^{(0)}}\otimes \ket{\xi^{(1)}} $ is the  initial state of the  apparatus $ \ha:= \h\sub{\aa}^{(0)}\otimes \h\sub{\aa}^{(1)}$, composed of both probes.

Now we may perform the TPM protocol on the total compound system so as to determine the total work distribution given the total unitary operator in \eq{eq:total-unitary-system-apparatus}.  For this  to be consistent with the original TPM protocol on the system alone, however, we require that when averaging out  the energy measurements performed on the apparatus, we must obtain the probability distribution given in \eq{eq:TPM-probability}, for all system states $\rho \in \s(\h)$ and system-only unitary operators $V$. In order for this to be satisfied, we demand that $\ket{\xi^{(j)}}$ be an eigenstate of the probe Hamiltonian $H\sub{\aa}^{(j)}$, with eigenvalue $\lambda_0^{(j)}$. This will ensure that the initial ideal energy measurement of the apparatus will not disturb it, so that the unitary interaction between system and apparatus by the unitary operators $U^{(j)}$ will result in the same state transformation as discussed in \sect{sec:TPM-measurement-scheme}. 

Let us first note that, given the assumption that the apparatus is initially prepared in an energy eigenstate, and using \eq{eq:Luders-unitary} and \eq{eq:total-unitary-system-apparatus}, we can show that for all $\ket{\psi} \in \h$, 
\begin{align}\label{eq:TPM-premeasurement}
&V_\mathrm{tot}(\ket{\psi} \otimes \ket{\xi}) \nonumber \\
& \qquad = e^{-i \theta_1 \lambda_0^{(1)}} \sum_{m,n} P_n V P_m \ket{\psi} \otimes e^{-i \theta_0 H\sub{\aa}^{(0)}}\ket{\phi_m^{(0)}}\otimes \ket{\phi_n^{(1)}}.
\end{align}
 Here, we have used the fact that $\ket{\xi^{(1)}}$ is an energy eigenstate with eigenvalue $\lambda_0^{(1)}$ to infer that  the component of $V_\mathrm{tot}$ given by $e^{-i \theta_1 H\sub{\aa}^{(1)}}$ only induces a constant phase factor $e^{-i \theta_1 \lambda_0^{(1)}}$, which is not physically observable.   Using this, we may now examine the extended TPM protocol, which will  be as follows:

\begin{enumerate}[(i)]

\item At time $t = t_0'$, perform an ideal energy measurement on the  compound system $\h\otimes \ha$, initially prepared in the state $\rho \otimes \pr{\xi}$. Since  the probes $\h\sub{\aa}^{(j)}$ are initially prepared in an energy eigenstate  with energy eigenvalue $\lambda_0^{(j)}$, only  outcomes $(m, 0, 0)$ are observed with non-zero probability, which result in  the compound system being prepared in the (unnormalized) state 
\begin{align}
&P_m \rho P_m \otimes Q_{0}^{(0)} \pr{\xi^{(0)}} Q_{0}^{(0)} \otimes Q_{0}^{(1)} \pr{\xi^{(1)}} Q_{0}^{(1)} \nonumber \\
& \qquad = P_m \rho P_m \otimes  \pr{\xi^{(0)}}  \otimes  \pr{\xi^{(1)}}. 
\end{align}

\item Between time $t_0'$ and $t_1'$,  let the compound system evolve according to the total unitary operator $V_\mathrm{tot}$ defined in \eq{eq:total-unitary-system-apparatus} and \eq{eq:TPM-premeasurement}. This prepares the (unnormalized) state
\begin{align}
& V_\mathrm{tot}(P_m \rho P_m \otimes  \pr{\xi^{(0)}}  \otimes  \pr{\xi^{(1)}}) V_\mathrm{tot}^\dagger \nonumber \\
& = \sum_{n,n'} P_n V P_m \rho P_{m} V^\dagger P_{n'} \otimes \tilde{P}[\phi_m^{(0)}] \otimes |\phi_n^{(1)}\>\<\phi_{n'}^{(1)}|, 
\end{align}
where $ \tilde{P}[\phi_m^{(0)}] := e^{-i \theta_0 H\sub{\aa}^{(0)}}\pr{\phi_m^{(0)}} e^{i \theta_0 H\sub{\aa}^{(0)}}$. 

\item At time $t = t_1'$, perform an ideal energy measurement on the  compound system $\h\otimes \ha$. Given the outcomes $(n, \mu, \nu )$, this prepares the (unnormalized) state
\begin{align}\label{eq:final-total-TPM-state}
P_n V P_m \rho P_{m} V^\dagger P_{n} \otimes Q_{\mu}^{(0)}\tilde{P}[\phi_m^{(0)}] Q_{\mu}^{(0)}\otimes Q_{\nu}^{(1)}\pr{\phi_n^{(1)}} Q_{\nu}^{(1)}.
\end{align}

\end{enumerate}

The full sequence of measurement outcomes is thus $X := (x, (0, \mu), (0, \nu))$, where $x := (m,n)$ is the sequence of outcomes for the system, while $(0, \mu)$ and $(0, \nu)$ are the sequences of outcomes for probes $\h\sub{\aa}^{(0)}$ and $\h\sub{\aa}^{(1)}$, respectively. The sequence $X$  corresponds with the total work done $\ww(X) := w(x) + w\sub{\aa}^{(0)}(\mu) +  w\sub{\aa}^{(1)}(\nu)$, where $w(x) := \epsilon_n - \epsilon_m$ is the contribution to the total work from the system, while $w\sub{\aa}^{(j)}(\mu) := \lambda_\mu^{(j)} - \lambda_0^{(j)}$ is the contribution to the total work from probe $\h\sub{\aa}^{(j)}$. The probability of observing sequence $X$, meanwhile, is given by the trace of the final unnormalized state \eq{eq:final-total-TPM-state}, which is 
\begin{align}\label{eq:total-work-probability-1}
p^{V_\mathrm{tot}}_{\rho, \xi}(X) &:=  p^V_\rho(x) \tr[Q_{\mu}^{(0)} \pr{\phi_m^{(0)}}] \tr[ Q_{\nu}^{(1)}\pr{\phi_n^{(1)}}],
\end{align}
where we recall that $p^V_\rho(x)$ is defined in \eq{eq:TPM-probability}. Note that here, we have used the fact that $Q_{\mu}^{(0)}$ is a spectral projection of $H\sub{\aa}^{(0)}$ to infer that $\tr[Q_{\mu}^{(0)}\tilde{P}[\phi_m^{(0)}]] = \tr[ Q_{\mu}^{(0)}e^{-i \theta_0 H\sub{\aa}^{(0)}}\pr{\phi_m^{(0)}} e^{i \theta_0 H\sub{\aa}^{(0)}}] = \tr[Q_{\mu}^{(0)} \pr{\phi_m^{(0)}}]$. 

Given that $\sum_\mu Q_{\mu}^{(0)} = \sum_\nu Q_{\nu}^{(1)} = \one$,   the marginal probability distribution for the system-only work will read as
\begin{align}
\sum_{\mu, \nu}p^{V_\mathrm{tot}}_{\rho, \xi}(X) &=  p^V_\rho(x) \tr[ \pr{\phi_m^{(0)}}] \tr[ \pr{\phi_n^{(1)}}]= p^V_\rho(x),
\end{align}
and so the extended TPM protocol on the compound of system plus apparatus is self-consistent.

\subsection{Satisfying the unmeasured first law for the total work}\label{sec:average-first-law}

Since the apparatus is initially prepared in an energy eigenstate, the average total work, for the total unitary process discussed in the previous section, clearly satisfies 
\begin{align}\label{eq:average-total-work}
\avg{\ww}^{V_\mathrm{tot}}_{\rho, \xi} &:= \sum_X p^{V_\mathrm{tot}}_{\rho, \xi}(X) \ww(X), \nonumber \\
& = \tr[(V_\mathrm{tot}^\dagger H_\mathrm{tot} V_\mathrm{tot} - H_\mathrm{tot}) \ii^L_\mm(\rho) \otimes \pr{\xi}] .
\end{align}
To see this, simply compare with \eq{eq:TPM-average-work}. As before, if $\rho$ does not commute with the Hamiltonian $H$, the total work is not guaranteed to satisfy the unmeasured first law, i.e., it is possible for some $\rho$ and $V$ to have  \eq{eq:average-total-work} differ from \eq{eq:total-unmeasured-work} (note that both the apparatus state $\ket{\xi}$, and the contribution to $V_\mathrm{tot}$ from the system-apparatus coupling, i.e., the unitaries $U^{(j)}$, are always fixed). However, as we shall show below, if additionally  the probe Hamiltonians $H\sub{\aa}^{(j)}$ are equivalent to the pointer observables $Z^{(j)}$, i.e., if we have 
\begin{align}\label{eq:pointer-observable-Hamiltonian-cond}
H\sub{\aa}^{(j)} = \sum_{m } \lambda_m^{(j)} Z_m^{(j)}, 
\end{align}
the unmeasured first law is guaranteed to be satisfied for the total work. 

Let us re-examine the TPM protocol on the compound of system plus apparatus once more in detail, this time assuming that \eq{eq:pointer-observable-Hamiltonian-cond} holds:

\begin{enumerate}[(i)]

\item At time $t = t_0'$ perform an ideal energy measurement on the  compound system $\h\otimes \ha$, initially prepared in the state $\rho \otimes \pr{\xi}$. As before, we assume that $\ket{\xi^{(j)}}$ are energy eigenstates, with eigenvalues $\lambda_0^{(j)}$, and hence only the outcomes $(m, 0, 0)$ are observed with non-zero probability, resulting  in the compound system to be prepared in the (unnormalized) state 
\begin{align}
&P_m \rho P_m \otimes Z_{0}^{(0)} \pr{\xi^{(0)}} Z_{0}^{(0)} \otimes Z_{0}^{(1)} \pr{\xi^{(1)}} Z_{0}^{(1)} \nonumber \\
& \quad = P_m \rho P_m \otimes  \pr{\xi^{(0)}}  \otimes  \pr{\xi^{(1)}}. 
\end{align}

\item Between time $t_0'$ and $t_1'$,  let the compound system evolve according to the total unitary operator $V_\mathrm{tot}$ defined in \eq{eq:total-unitary-system-apparatus} and \eq{eq:TPM-premeasurement}. This prepares the (unnormalized) state
\begin{align}
& V_\mathrm{tot}(P_m \rho P_m \otimes  \pr{\xi^{(0)}}  \otimes  \pr{\xi^{(1)}}) V_\mathrm{tot}^\dagger \nonumber \\
& \quad  = \sum_{n,n'} P_n V P_m \rho P_{m} V^\dagger P_{n'} \otimes \pr{\phi_m^{(0)}} \otimes |\phi_n^{(1)}\>\<\phi_{n'}^{(1)}|. 
\end{align}
Note that since $\ket{\phi_m^{(0)}}$ are eigenvalue-1 eigenstates of the projection operators $Z_m^{(0)}$, which clearly commute with the Hamiltonian, it follows that  $ e^{-i \theta_0 H\sub{\aa}^{(0)}}\pr{\phi_m^{(0)}} e^{i \theta_0 H\sub{\aa}^{(0)}} = \pr{\phi_m^{(0)}}$. 

\item At time $t=t_1'$, perform an ideal energy measurement on the  compound system $\h\otimes \ha$. Given outcomes $(n, n', n'' )$, this prepares the (unnormalized) state
\begin{align}\label{eq:final-total-TPM-state-pointer-Hamiltonian}
&P_n V P_m \rho P_{m} V^\dagger P_{n} \otimes Z_{n'}^{(0)}\pr{\phi_m^{(0)}} Z_{n'}^{(0)}\otimes Z_{n''}^{(1)}\pr{\phi_n^{(1)}} Z_{n''}^{(1)} \nonumber \\
& \quad = \delta_{m,n'} \delta_{n, n''} P_n V P_m \rho P_{m} V^\dagger P_{n} \otimes \pr{\phi_m^{(0)}}\otimes \pr{\phi_n^{(1)}},
\end{align}
where the final line follows from the fact that $Z_n^{(j)}\ket{\phi_m^{(j)}} = \delta_{m,n}\ket{\phi_m^{(j)}}$.
\end{enumerate}

Equation \eqref{eq:final-total-TPM-state-pointer-Hamiltonian} implies that the only sequences of energy measurement outcomes that are observed with non-zero probability are $X := (x, (0,m), (0,n))$, where we recall that $x := (m,n)$. In other words, the energy transitions of the apparatus fully determine the energy transitions of the system, and vice versa. As such, let us remove some of the redundancy and write $X := ((0,0) ,x)$, where $(0,0)$ denotes the energy measurement outcomes on the apparatus at time $t_0'$, and $x= (m,n)$ denotes both the energy measurement outcomes on the apparatus at time $t_1'$, as well as the sequence of energy measurement outcomes on the system at times $t_0', t_1'$. The total work done given the sequence $X$ is thus $\ww(X) := w(x) + w\sub{\aa}^{(0)}(m) +  w\sub{\aa}^{(1)}(n)$, where $w\sub{\aa}^{(j)}(m) := \lambda^{(j)}_m - \lambda^{(j)}_0$,  with the probability 
\begin{align}\label{eq:total-work-probability}
p^{V_\mathrm{tot}}_{\rho, \xi}(X) &=  p^V_\rho(x). 
\end{align}
Note that this is equivalent to \eq{eq:total-work-probability-1} when we replace $Q_\mu^{(j)}$ with $Z_{m}^{(j)}$, which gives $\tr[Z_{m}^{(j)} \pr{\phi_m^{(j)}}] = 1$.

As shown in \app{app:average-first-law}, the average total work will now read as
\begin{align}\label{eq:average-first-law-total-work}
\avg{\ww}^{V_\mathrm{tot}}_{\rho, \xi} &:= \sum_{X} p^{V_\mathrm{tot}}_{\rho, \xi}(X) \ww(X)  , \nonumber \\
 &= \tr[(V_\mathrm{tot}^\dagger H_\mathrm{tot} V_\mathrm{tot} - H_\mathrm{tot}) (\rho \otimes \pr{\xi})]
\end{align}
for all $\rho \in \s(\h)$ and $V$. As such, comparing with \eq{eq:total-unmeasured-work} we see that so long as the apparatus is initially prepared in an energy eigenstate, and \eq{eq:pointer-observable-Hamiltonian-cond} is satisfied, then not only will the TPM protocol for the total work be self-consistent, but the  total work will always satisfy the unmeasured first law, even for initial system states $\rho$ that do not commute with the Hamiltonian. As a final observation, note that \eq{eq:average-total-work} must be equivalent to \eq{eq:average-first-law-total-work} when \eq{eq:pointer-observable-Hamiltonian-cond} is satisfied. Consequently,  in such a case  the following equality holds:
\begin{align}
&\tr[(V_\mathrm{tot}^\dagger H_\mathrm{tot} V_\mathrm{tot} - H_\mathrm{tot}) (\rho \otimes \pr{\xi})] \nonumber \\
& \quad = \tr[(V_\mathrm{tot}^\dagger H_\mathrm{tot} V_\mathrm{tot} - H_\mathrm{tot}) (\ii^L_\mm(\rho) \otimes \pr{\xi})]
\end{align}
for all $\rho \in \s(\h)$ and $V$. This is a consequence of the strong repeatability of ideal energy measurements \cite{Busch1995}, which implies that directly performing ideal energy measurements on the system is redundant; the structure of the unitary operators $U^{(j)}$, together with the fact that we  measure the probes by the pointer observables $Z^{(j)}$, ensures that the system automatically undergoes an ideal energy measurement.

\subsection{ Necessary and sufficient conditions for the  total work distribution to be equal to the system-only work distribution}

If the apparatus is initially prepared in an energy eigenstate, and  the apparatus probe Hamiltonians  are equivalent to  the pointer observable, the probability distribution for the total work $\ww$, given an initial  total state $\rho \otimes \pr{\xi}$ and total unitary operator $ V_\mathrm{tot}$, is  
\begin{align}\label{eq:distribution-total-work}
p_{\rho,\xi}^{ V_\mathrm{tot}}(\ww) &:= \sum_{X} \delta(\ww - \ww(X)) p_{\rho,\xi}^{ V_\mathrm{tot}}(X), \nonumber \\
& \equiv \sum_x \delta(\ww - (w(x) + w\sub{\aa}(x) )) p_\rho^V(x),  
\end{align}
where we have used \eq{eq:total-work-probability}, together with the definition $w\sub{\aa}(x) := w\sub{\aa}^{(0)}(m) +  w\sub{\aa}^{(1)}(n)$. 

It is simple to see that, in general,  the total work probability distribution  \eq{eq:distribution-total-work} is different  to the system-only work probability distribution  \eq{eq:distribution-work}.  In order for these distributions to be the same, for all system states $\rho$ and system-only unitary operators $V$, we must have $w\sub{\aa}(x) = 0$ for all $x$ such that $p_\rho^V(x) >0$ for some $\rho$ and $V$. This ensures that  for all $\rho$ and $V$, 
\begin{align}
\sum_{x} \delta(w - w(x)) p_\rho^V(x) = \sum_{x} \delta(w - \ww(X)) p_\rho^V(x). 
\end{align}
Recall that $w\sub{\aa}^{(j)}(m) := \lambda^{(j)}_m - \lambda^{(j)}_0 $, where  $\lambda^{(j)}_0$ is a fixed energy eigenvalue, and that $p_\rho^V(x) := \tr[P_m V^\dagger P_n V P_m \rho ]$. Consequently,  the condition $w\sub{\aa}(x)=0$ for all $x$ such that $p_\rho^V(x)>0$ for some $\rho$ and $V$  is equivalent to the condition $\lambda^{(j)}_m = \lambda^{(j)}_0$ for all $m$ such that $P_m >\zero$, i.e., if only a single degenerate energy subspace of the probe is involved during the measurement process.     Interestingly, we shall see that this condition is equivalent to the statement that the  unmeasured work given the unitary operator $U^{(j)}$  vanishes for all system states.

The unmeasured work, given the measurement unitary coupling between system and probe $\h\sub{\aa}^{(j)}$, is given as 
\begin{align}\label{eq:unmeasured-measurement-work}
W_\mathrm{meas}^{(j)}:= \tr[(U^{(j)\dagger }H_\mathrm{tot}^{(j)}U^{(j)} - H_\mathrm{tot}^{(j)})\rho \otimes \pr{\xi^{(j)}}],
\end{align}
where we define $H_\mathrm{tot}^{(j)} := H + H\sub{\aa}^{(j)}$ as the additive Hamiltonian of the composite system $\h\otimes \h\sub{\aa}^{(j)}$. This can equivalently be written as 
\begin{align}\label{eq:unmeasured-measurement-work-1}
W_\mathrm{meas}^{(j)} = \tr[\Gamma_{\xi^{(j)}}\left( U^{(j)\dagger }H_\mathrm{tot}^{(j)}U^{(j)} - H_\mathrm{tot}^{(j)}\right) \rho],
\end{align}
where $\Gamma_{\xi^{(j)}} : \lo(\h\otimes \h\sub{\aa}^{(j)}) \to \lo(\h)$ is the restriction map for $\ket{\xi^{(j)}}$, defined by the identity 
$ \tr[\Gamma_{\xi^{(j)}}(B) T] = \tr[B (T \otimes \pr{\xi^{(j)}})]$ for all $B \in \lo(\h\otimes \h\sub{\aa}^{(j)})$ and $T \in \trc(\h)$ \cite{Loveridge2017a}. Recalling that the  unitary operator  $ U^{(j)}$  always satisfies \eq{eq:Luders-unitary}, we thus have 
\begin{align}\label{eq:Luders-effective-conservation}
\Gamma_{\xi^{(j)}}\left( U^{(j)\dagger }H_\mathrm{tot}^{(j)}U^{(j)} - H_\mathrm{tot}^{(j)}\right) &= \sum_{m} w\sub{\aa}^{(j)}(m) P_m.
\end{align}
For a detailed proof, refer  to  \app{app:restriction-map}.  The right hand side of \eq{eq:Luders-effective-conservation} vanishes if for each $m$, either $P_m = \zero$, or $w\sub{\aa}^{(j)}(m) = 0$. Consequently,  $w\sub{\aa}^{(j)}(m)=0$ for all $m$ such that $P_m > \zero$ is necessary and sufficient for the left hand side of \eq{eq:Luders-effective-conservation} to vanish. But by \eq{eq:unmeasured-measurement-work-1} this implies that   $W_\mathrm{meas}^{(j)} =0$ for all  $\rho \in \s(\h)$; given that the apparatus is in the state $\ket{\xi^{(j)}}$, then irrespective of what state the system is prepared in,  the unmeasured work given the unitary operator $U^{(j)}$ will vanish. We refer to this as $U^{(j)}$ satisfying ``weak'' energy conservation,  which is a weaker condition than full energy conservation, i.e., $[H_\mathrm{tot}^{(j)}, U^{(j)}] = \zero$, which implies  that $W_\mathrm{meas}^{(j)} =0$ for all choices of the apparatus state $\ket{\xi^{(j)}}$. 

We note that while a fully degenerate probe Hamiltonian, $H\sub{\aa}^{(j)}= \lambda_0^{(j)} \one$, or a fully energy conserving  unitary,  $[H_\mathrm{tot}^{(j)}, U^{(j)}]=\zero$, are sufficient conditions for the total work distribution \eq{eq:distribution-total-work} to equal the system-only work distribution \eq{eq:distribution-work}, they are not necessary.  

To illustrate the first point, consider the system Hamiltonian $H = \epsilon_1 P_1 + \epsilon_2 P_2$,  where $P_1, P_2 > \zero$.  However, this is equivalent to   $H = \epsilon_1 P_1 + \epsilon_2 P_2 +  \epsilon_3 P_3$ such that $P_3 = \zero$. Therefore,  the ideal measurement of $H$ can be realized by the normal measurement scheme $(\ha, \ket{\xi}, U, Z)$,  with the three-valued pointer observable $Z := \{Z_1,Z_2,Z_3\}$, $Z_m >\zero$, and the  unitary operator $U$ which satisfies 
\begin{align}
U(\ket{\psi} \otimes \ket{\xi}) &= \sum_{m=1}^3 P_m \ket{\psi} \otimes \ket{\phi_m}
\end{align} 
for all $\ket{\psi} \in \h$, where $\ket{\phi_m}$ are eigenvalue-1 eigenstates of $Z_m$. Note that the term for $m=3$ vanishes, since $P_3 \ket{\psi} = \zero \ket{\psi} = 0$ for all $\ket{\psi}$; the apparatus is never taken to the state $\ket{\phi_3}$.  Let the apparatus have the Hamiltonian $H\sub{\aa} = \lambda (Z_1 + Z_2) + \lambda' Z_3$, where $\lambda \ne \lambda'$, so that $H\sub{\aa}$ is not fully degenerate. Notwithstanding, if $\ket{\xi}$ is in the support of  $Z_1+Z_2$, we still have $w\sub{\aa}(m) =  0$ for $m=1,2$, i.e., for all $m$ corresponding to $P_m > \zero$. As stated previously, it is only necessary that a single degenerate energy subspace of the apparatus be ``involved'' during the measurement process;  for all measurement outcomes that are observed, the state of the apparatus starts and ends in the support of $Z_1 + Z_2$.

To illustrate that full energy conservation by the  unitary  is also not necessary, consider the simple case where $\h\simeq \co^2$, with orthonormal basis $\{\ket{0}, \ket{1}\}$, and Hamiltonian $H = 
\epsilon |1\>\<1|$, $\epsilon > 0$. A normal measurement scheme for an ideal measurement of $H$ can be given as $(\ha, \ket{0}, U, Z )$,   where $\ha \simeq \co^2$,     $Z :=\{|0\>\<0|, |1\>\<1| \}$, and  
\begin{align}
U : 
\begin{cases}
\ket{m,0} \mapsto \ket{m,m} \\
\ket{m,1} \mapsto \ket{m \oplus_2 1,m}
\end{cases},
\end{align}
where $m=0,1$ and $\oplus_2$ denotes addition modulo 2. Note that only the transformation $\ket{m,0} \mapsto \ket{m,m}$ is ever utilized, since the apparatus is initially prepared in state $\ket{0}$.  If the apparatus Hamiltonian is fully degenerate, $H\sub{\aa} = \lambda \one$, then $U$ will satisfy weak energy conservation; given $H_\mathrm{tot} = H + H\sub{\aa}$, then for any $\ket{\psi} =\alpha |0\> + \beta |1\>$, we have $\<\psi,0| U^\dagger H_\mathrm{tot}  U |\psi,0\> = |\beta|^2\epsilon + \lambda = \<\psi,0| H_\mathrm{tot}  |\psi,0\> $.    However, $[U, H_\mathrm{tot}] \ne \zero$, since $U \ket{1,1} = \ket{0,1}$, and hence $\<1,1| U^\dagger H_\mathrm{tot} U|1,1\> = \lambda \ne \<1,1| H_\mathrm{tot} |1,1\> = \epsilon + \lambda$.

\section{ Conclusions}

A definition for work which relies on measurements is self-consistent if it can  account for the contribution to  work by the measurement process itself, at least in principle. More precisely, for self-consistency we demand that the marginal of the total work distribution for system and measurement apparatus, obtained by ignoring the apparatus degrees of freedom,  recovers the original work distribution for the system alone. In the case of the two-point energy measurement (TPM) protocol, we have shown that this is possible so long as the measurement apparatus is initially prepared in an energy eigenstate. Furthermore, if the apparatus Hamiltonian is chosen to be equivalent to the pointer observable, then the total work distribution will always satisfy the unmeasured first law; the average total work will equal the change in average energy given the total unitary evolution. This is  a consequence of the strong repeatability of ideal energy measurements, which implies that directly performing energy measurements on the system is redundant.  Finally, we have shown that the total work distribution will be identical to the system-only work distribution if and only if the unmeasured work, given the unitary interaction between system and apparatus, vanishes for all system states. Extending the present framework of analysis to other definitions of work remain as open questions for further research.

\begin{acknowledgments} The author  acknowledges support from  the Slovak Academy of Sciences   under MoRePro project OPEQ (19MRP0027), as 
well as projects OPTIQUTE (APVV-18-0518) and HOQIT (VEGA 2/0161/19).
\end{acknowledgments}


\bibliographystyle{apsrev4-1}
\bibliography{Projects-TPM-consistency}

%
\appendix

\widetext

\section{Proof of \eq{eq:average-first-law-total-work}}\label{app:average-first-law}

Let us introduce the operation (completely positive and trace non-increasing map) $\jj_{x',x} : \trc(\h\otimes \ha) \to \trc(\h\otimes \ha)$, defined as 
\begin{align}\label{Instrument-total-TPM}
\jj_{x',x}(\cdot) :=(\one \otimes Z_x) V_\mathrm{tot}(\one \otimes Z_{x'}) (\cdot)  (\one \otimes Z_{x'}) V_\mathrm{tot}^\dagger (\one \otimes Z_{x}).
\end{align}
Here,  $x:= (m,n)$ and $x' := (m', n')$, so that $Z_x := Z_m^{(0)}\otimes Z_n^{(1)}$ and $Z_{x'} := Z_{m'}^{(0)}\otimes Z_{n'}^{(1)}$, and $V_\mathrm{tot}$ is defined in \eq{eq:total-unitary-system-apparatus}. First, let us show that the state transformation given the TPM protocol can be fully described by the operation $\jj_{x',x}$. Denoting $(0,0) \equiv 0$,  we find that for any $T \in \trc(\h)$, $ \pr{\xi} :=  \pr{\xi^{(0)}}\otimes  \pr{\xi^{(1)}}$, and $x$,  the following:
\begin{align}\label{Instrument-total-TPM-state}
\jj_{0,x}(T \otimes \pr{\xi}) &=  (\one \otimes Z_x) V_\mathrm{tot}(\one \otimes Z_{0}) (T \otimes \pr{\xi})  (\one \otimes Z_0) V_\mathrm{tot}^\dagger (\one \otimes Z_{x}) , \nonumber \\
& = (\one \otimes Z_x) V_\mathrm{tot} (T \otimes \pr{\xi})  V_\mathrm{tot}^\dagger (\one \otimes Z_{x}), \nonumber \\
& = \sum_{m',m'',n',n''} (\one \otimes Z_x)  \left(P_{n'} V P_{m'} T P_{m''} V^\dagger P_{n''} \otimes |\phi_{m'}^{(0)}\>\<\phi_{m''}^{(0)}|\otimes |\phi_{n'}^{(1)}\>\<\phi_{n''}^{(1)}|\right)  (\one \otimes Z_{x}), \nonumber \\
& = P_n V P_m T P_{m} V^\dagger P_{n} \otimes \pr{\phi_m^{(0)}}\otimes \pr{\phi_n^{(1)}}.
\end{align}
In the second line, we have used the fact that $\ket{\xi}$ is an eigenvalue-1 eigenstate of $Z_0 = Z_0^{(0)} \otimes Z_0^{(1)}$. In the third line, we have used  \eq{eq:total-unitary-system-apparatus} and  \eq{eq:TPM-premeasurement}. In the final line, we have used the fact that $\ket{\phi^{(j)}_m}$ are eigenvalue-1  eigenstates of $Z_m^{(j)}$.   Similarly, it is easy to show that $\jj_{x',x}(T \otimes \pr{\xi}) = \zero$ for all $x' \ne (0,0)$. 

Recall that, given the sequence $X := (0, x)$, the TPM work done is 
\begin{align}\label{total-work-1}
\ww(X) &:= w(x) + w\sub{\aa}^{(0)}(m) + w\sub{\aa}^{(1)}(n), \nonumber \\
& = (\epsilon_n +\lambda^{(0)}_m + \lambda^{(1)}_n ) - (\epsilon_m + \lambda_0^{(0)} + \lambda_0^{(1)}).
\end{align}

 Using \eq{Instrument-total-TPM-state},  recalling that $H_\mathrm{tot} = H + H\sub{\aa}^{(0)} + H\sub{\aa}^{(1)}$, and that $\ket{\xi}$ and $\ket{\phi_m^{(j)}}$ are eigenstates of $H\sub{\aa}^{(0)} + H\sub{\aa}^{(1)}$,  we may verify that 
 \begin{align}
 \tr[H_\mathrm{tot} \jj_{0,x}(\rho \otimes \pr{\xi})] &= (\epsilon_n +\lambda^{(0)}_m + \lambda^{(1)}_n )\tr[\jj_{0,x}(\rho \otimes \pr{\xi})], \nonumber \\
 \tr[ \jj_{0,x}(H_\mathrm{tot}(\rho \otimes \pr{\xi}) )] & = (\epsilon_m + \lambda_0^{(0)} + \lambda_0^{(1)})\tr[ \jj_{0,x}(\rho \otimes \pr{\xi})].
 \end{align}
 Consequently, we may express \eq{total-work-1} as
\begin{align}
\ww(X) &=  \frac{\tr[H_\mathrm{tot}\jj_{0,x} (\rho \otimes \pr{\xi})]}{\tr[\jj_{0,x}(\rho \otimes \pr{\xi})]}   - \frac{\tr[\jj_{0,x}(H_\mathrm{tot} \rho \otimes \pr{\xi} )]}{\tr[\jj_{0,x}(\rho \otimes \pr{\xi})]}.
\end{align}
Recalling that $  p^{V_\mathrm{tot}}_{\rho, \xi}(X) = \tr[\jj_{0,x}(\rho \otimes \pr{\xi})]$,  we may therefore write the average work as 
\begin{align}
\avg{\ww}^{V_\mathrm{tot}}_{\rho, \xi} &:= \sum_{X} p^{V_\mathrm{tot}}_{\rho, \xi}(X) \ww(X), \nonumber \\
& = \sum_{x',x} \tr[H_\mathrm{tot}\jj_{x',x} (\rho \otimes \pr{\xi})] - \sum_{x',x} \tr[\jj_{x',x}(H_\mathrm{tot} \rho \otimes \pr{\xi} )].
\end{align}
Noting that $\sum_{x',x} \jj_{x', x}$ is a trace-preserving operation, it follows that 
\begin{align}
\sum_{x',x} \tr[\jj_{x',x}(H_\mathrm{tot} \rho \otimes \pr{\xi} )] = \tr[H_\mathrm{tot}( \rho \otimes \pr{\xi})].
\end{align} 
Similarly, noting that $\sum_{x'}Z_{x'} H_\mathrm{tot} Z_{x'} = H_\mathrm{tot}$, while $\sum_{x} Z_x \pr{\xi} Z_x = \pr{\xi}$, we have
\begin{align}
 \sum_{x',x} \tr[H_\mathrm{tot}\jj_{x',x} (\rho \otimes \pr{\xi})] = \tr[H_\mathrm{tot} V_\mathrm{tot}(\rho \otimes \pr{\xi})V_\mathrm{tot}^\dagger].
\end{align}
Therefore, the average total work reads
\begin{align}
\avg{\ww}^{V_\mathrm{tot}}_{\rho, \xi} &= \tr[(V_\mathrm{tot}^\dagger H_\mathrm{tot} V_\mathrm{tot} - H_\mathrm{tot}) (\rho \otimes \pr{\xi})]
\end{align}
for all $\rho \in \s(\h)$ and $V$.

\section{Proof of \eq{eq:Luders-effective-conservation}}\label{app:restriction-map}

Let $H = \sum_m \epsilon_m P_m$ be the Hamiltonian of system $\h$, and $H\sub{\aa} = \sum_m \lambda_m Z_m$ the Hamiltonian of system $\ha$, such that $H_\mathrm{tot} = H + H\sub{\aa}$ is the total, additive Hamiltonian of the compound system $\h\otimes \ha$. Moreover, let $\ket{\xi} \in \ha$ be a unit vector which is an eigenstate of $H\sub{\aa}$ with eigenvalue $\lambda_0$. Finally, let $U$ be a unitary operator on $\h\otimes \ha$ such that, for all $\ket{\psi} \in \h$, 
\begin{align}\label{eq:app-unitary}
\ket{U(\psi \otimes \xi)} = \sum_m \ket{P_m\psi \otimes \phi_m},
\end{align}
 where $\ket{\psi \otimes \xi} \equiv \ket{\psi}\otimes \ket{\xi}$, $\ket{P_m \psi} \equiv P_m\ket{\psi}$, and $\ket{\phi_m}$ are  eigenstates of $H\sub{\aa}$ with eigenvalue $\lambda_m$. It follows that for all $\ket{\psi} \in \h$, we have 
\begin{align}\label{eq:app-inner-prod-1}
\<\psi \otimes \xi| H_\mathrm{tot}|\psi \otimes \xi\> &= \<\psi| H |\psi\>\<\xi|\xi\> + \<\psi|\psi\>\<\xi| H\sub{\aa} |\xi\>, \nonumber \\ 
& = \<\psi| H |\psi\> +  \<\psi|\psi\> \lambda_0, \nonumber \\
&=  \<\psi|\big(H + \lambda_0\one\big) |\psi\>. 
\end{align}
In the first line we use the additivity of $H_\mathrm{tot}$, in the second line we use the fact that $\<\xi|\xi\>=1$ and $\<\xi| H\sub{\aa} |\xi\> = \lambda_0$, and in the final line we use the fact that $\<\psi|\psi\> \lambda_0 = \<\psi| \lambda_0 \one |\psi\>$. Similarly, for all $\ket{\psi}\in \h$ we have
 \begin{align}\label{eq:app-inner-prod-2}
\<\psi \otimes \xi|U^{\dagger} H_\mathrm{tot} U|\psi \otimes \xi\> & = \<U(\psi \otimes \xi)| H_\mathrm{tot} |U(\psi \otimes \xi)\>, \nonumber \\
&=  \sum_{m,n}\<P_m \psi \otimes \phi_m|H_\mathrm{tot} |P_n \psi \otimes \phi_n\>, \nonumber \\
& = \sum_{m,n}\< \psi|P_m H  P_n|  \psi\>\<\phi_m| \phi_n\> + \<\psi| P_m P_n |\psi\>\<\phi_m| H\sub{\aa} |\phi_n\> , \nonumber \\
& = \sum_m \<\psi| P_m H P_m |\psi\> +  \<\psi| P_m  |\psi\>\lambda_m, \nonumber \\
& = \<\psi| \big(H + \sum_{m} \lambda_m P_m \big)|\psi\>.
\end{align}
In the first line we use the definition of the adjoint of a unitary operator $U$, in the second line we use \eq{eq:app-unitary}, in the third line we use the additivity of $H_\mathrm{tot}$ and the fact that the projection operators $P_m$ are self-adjoint, in the fourth line we use $\<\phi_m| \phi_n\> = \delta_{m,n}$ together with  $P_m P_n = \delta_{m,n}P_m$ and $\<\phi_m| H\sub{\aa} |\phi_m\> = \lambda_m$, and in the final line we use the fact that $\sum_m P_m H P_m = H$.   

Combining \eq{eq:app-inner-prod-1} with \eq{eq:app-inner-prod-2} implies that, for all $\ket{\psi} \in \h$, we have 
\begin{align}
\<\psi \otimes \xi|(U^{\dagger} H_\mathrm{tot} U - H_\mathrm{tot})|\psi \otimes \xi\> &= \<\psi| \big(\sum_{m} \lambda_m P_m - \lambda_0 \one \big)|\psi\>, \nonumber \\
& = \<\psi| \big(\sum_{m} (\lambda_m - \lambda_0) P_m  \big)|\psi\>,
\end{align}
where in the final line we use the fact that $\sum_m P_m = \one$. From the above equation, it follows that 
\begin{align}
\Gamma_\xi(U^{\dagger} H_\mathrm{tot} U - H_\mathrm{tot}) = \sum_{m} (\lambda_m - \lambda_0) P_m,
\end{align}
where $\Gamma_\xi : \lo(\h\otimes \ha) \to \lo(\h)$ is the restriction map for $\ket{\xi} \in \ha$ defined as $\tr[\Gamma_\xi(B) T] = \tr[B(T \otimes \pr{\xi})]$ for all $B \in \lo(\h\otimes \ha)$ and $T \in \trc(\h)$.

%
%
%
%
%
%
%

\end{document}